\documentclass[a4paper,12pt]{spieman}  
\usepackage{amsmath,amsfonts,amssymb}
\usepackage{graphicx} 
\usepackage{multirow}
\usepackage{setspace}
\usepackage{tocloft}
\usepackage{multicol}
\usepackage[left]{lineno} 

\newcommand{\tus}{a}
\newcommand{\ut}{b}
\newcommand{\rceu}{c}
\newcommand{\ipmu}{d}
\newcommand{\jaxa}{e}
\newcommand{\isas}{f}
\newcommand{\hiroshima}{g}
\newcommand{\hasc}{h}
\newcommand{\tksc}{i}
\newcommand{\eotvos}{j}
\newcommand{\mta}{k}
\newcommand{\kmi}{l}
\newcommand{\kagura}{m}
\newcommand{\slac}{n}
\newcommand{\chubu}{o}
\newcommand{\kipac}{p}
\newcommand{\rikkyo}{q}
\newcommand{\hakubi}{r}
\newcommand{\waseda}{s}
\newcommand{\cnrs}{t}
\newcommand{\cea}{u}
\newcommand{\riken}{v}
\newcommand{\yamagata}{w}
\newcommand{\osaka}{x}
\newcommand{\prcfs}{y}
\newcommand{\isee}{z}
\newcommand{\kyoto}{aa}
\newcommand{\saitama}{ab}
\newcommand{\shizuoka}{ac}
\newcommand{\ai}{ad}
\newcommand{\tit}{ae}

\title{{Origin of the in-orbit instrumental background of\\ the Hard X-ray Imager onboard Hitomi}}

\author[\tus,*]{Kouichi~Hagino} 
\author[\ut,\rceu,\ipmu]{Hirokazu~Odaka}
\author[\jaxa]{Goro~Sato}
\author[\ut,\isas]{Tamotsu~Sato}
\author[\ut]{Hiromasa~Suzuki}
\author[\hiroshima,\hasc]{Tsunefumi~Mizuno}
\author[\tksc]{Madoka~Kawaharada}
\author[\eotvos,\mta,\hiroshima]{Masanori~Ohno} 
\author[\kmi]{Kazuhiro~Nakazawa}
\author[\kagura]{Shogo~B.~Kobayashi}
\author[\ut]{Hiroaki~Murakami}
\author[\ut]{Katsuma~Miyake} 
\author[\slac]{Makoto~Asai}
\author[\chubu]{Tatsumi~Koi}
\author[\kipac,\slac]{Greg~Madejski}
\author[\rikkyo]{Shinya~Saito}
\author[\slac]{Dennis~H.~Wright} 
\author[\hakubi]{Teruaki~Enoto} 
\author[\hiroshima]{Yasushi~Fukazawa} 
\author[\isas]{Katsuhiro~Hayashi}
\author[\waseda]{Jun~Kataoka}
\author[\hiroshima]{Junichiro~Katsuta} 
\author[\isas]{Motohide~Kokubun}
\author[\cnrs,\cea]{Philippe~Laurent} 
\author[\cnrs,\cea]{Fran\c{c}ois~Lebrun} 
\author[\cea]{Olivier~Limousin}
\author[\cea]{Daniel~Maier}
\author[\ipmu,\ut,\riken]{Kazuo~Makishima} 
\author[\isas]{Kunishiro~Mori}
\author[\yamagata]{Takeshi~Nakamori} 
\author[\riken]{Toshio~Nakano}
\author[\osaka,\prcfs]{Hirofumi~Noda}
\author[\isas]{Masayuki~Ohta}
\author[\isas]{Rie~Sato}
\author[\isee]{Hiroyasu~Tajima} 
\author[\hiroshima]{Hiromitsu~Takahashi} 
\author[\ipmu,\ut]{Tadayuki~Takahashi}
\author[\ipmu]{Shin'ichiro~Takeda}
\author[\kyoto]{Takaaki~Tanaka}
\author[\saitama]{Yukikatsu~Terada} 
\author[\shizuoka]{Hideki~Uchiyama}
\author[\ai,\rikkyo]{Yasunobu~Uchiyama} 
\author[\isas,\ipmu]{Shin~Watanabe}
\author[\isee]{Kazutaka~Yamaoka}
\author[\tit]{Yoichi~Yatsu}
\author[\riken]{Takayuki~Yuasa}

\affil[\tus]{Department of Physics, Tokyo University of Science, 2641 Yamazaki, Noda, Chiba, 278-8510, Japan}
\affil[\ut]{Department of Physics, The University of Tokyo, 7-3-1 Hongo, Bunkyo-ku, Tokyo 113-0033, Japan}
\affil[\rceu]{Research Center for the Early Universe, School of Science, The University of Tokyo, 7-3-1 Hongo, Bunkyo-ku, Tokyo 113-0033, Japan}
\affil[\ipmu]{Kavli IPMU (WPI), UTIAS, The University of Tokyo, 5-1-5 Kashiwanoha, Kashiwa, Chiba 277-8583, Japan}
\affil[\jaxa]{Japan Aerospace Exploration Agency, 4-6 Kandasurugadai, Chiyoda-ku, Tokyo 101-8008, Japan}
\affil[\isas]{Institute of Space and Astronautical Science, Japan Aerospace Exploration Agency, 3-1-1 Yoshino-dai, Chuo-ku, Sagamihara, Kanagawa 252-5210, Japan}
\affil[\hiroshima]{School of Science, Hiroshima University, 1-3-1 Kagamiyama, Higashi-Hiroshima 739-8526, Japan}
\affil[\hasc]{Hiroshima Astrophysical Science Center, Hiroshima University, Higashi-Hiroshima, Hiroshima 739-8526, Japan}
\affil[\tksc]{Space Technology Directorate I, Japan Aerospace Exploration Agency, 2-1-1 Sengen, Tsukuba, Ibaraki 305-8505, Japan}
\affil[\eotvos]{E\"otv\"os University, Institute of Physics, P\'azm\'any P\'eter s\'et\'any 1/A, Budapest, 1117, Hungary}
\affil[\mta]{MTA-E\"otv\"os University Lend\"ulet Hot Universe and Astrophysics  Research Group, P\'azm\'any P\'eter s\'et\'any 1/A, Budapest, 1117, Hungary}
\affil[\kmi]{Kobayashi-Maskawa Institute for the Origin of Particles and the Universe, Nagoya University, Nagoya, Japan}
\affil[\kagura]{Department of Physics, Tokyo University of Science, 1-3 Kagurazaka, Shinjuku-ku, Tokyo 162-8601, Japan}
\affil[\slac]{SLAC National Accelerator Laboratory, 2575 Sand Hill Road, Menlo Park, CA 94025, USA}
\affil[\chubu]{College of Engineering, Chubu University, 1200 Matsumoto-cho, Kasugai, Aichi 487-8501, Japan}
\affil[\kipac]{Kavli Institute for Particle Astrophysics and Cosmology, Stanford University, 452 Lomita Mall, Stanford, CA 94305, USA}
\affil[\rikkyo]{Department of Physics, Rikkyo University, 3-34-1 Nishi-Ikebukuro, Toshima-ku, Tokyo 171-8501, Japan}
\affil[\hakubi]{Extreme Natural Phenomena RIKEN Hakubi Research Team, RIKEN Cluster for Pioneering Research, 2-1 Hirosawa, Wako, Saitama 351-0198, Japan}
\affil[\waseda]{Research Institute for Science and Engineering, Waseda University, 3-4-1 Ohkubo, Shinjuku, Tokyo, 169-8555, Japan}
\affil[\cnrs]{Universit\'e de Paris, CNRS, Astroparticule et Cosmologie, F-75013 Paris, France}
\affil[\cea]{CEA/DRF/IRFU/DAP - AIM, CEA/CNRS/Universit\'e Paris-Saclay, Universit\'e de Paris, F-91191 Gif-sur-Yvette, France}
\affil[\riken]{High Energy Astrophysics Laboratory, Nishina Center, RIKEN, Saitama, Japan}
\affil[\yamagata]{Faculty of Science, Yamagata University, 1-4-12 Kojirakawa-machi, Yamagata, Yamagata 990-8560, Japan}
\affil[\osaka]{Department of Earth and Space Science, Osaka University, 1-1 Machikaneyama-cho, Toyonaka, Osaka 560-0043, Japan}
\affil[\prcfs]{Project Research Center for Fundamental Sciences, Osaka University, 1-1 Machikaneyama-cho, Toyonaka, Osaka 560-0043, Japan}
\affil[\isee]{Institute for Space-Earth Environmental Research, Nagoya University, Furo-cho, Chikusa-ku, Nagoya, Aichi 464-8601, Japan}
\affil[\kyoto]{Department of Physics, Kyoto University, Kitashirakawa-Oiwake-Cho, Sakyo, Kyoto 606-8502, Japan}
\affil[\saitama]{Graduate School of Science and Engineering, Physics, Saitama University, 255 Shimo-Okubo, Sakura-ku, Saitama, 338-8570, Japan}
\affil[\shizuoka]{Faculty of Education, Shizuoka University, 836 Ohya, Suruga-ku, Shizuoka 422-8529, Japan}
\affil[\ai]{Graduate School of Artificial Intelligence and Science, Rikkyo University 3-34-1 Nishi-Ikebukuro, Toshima-ku, Tokyo 171-8501, Japan}
\affil[\tit]{Department of Physics, Tokyo Institute of Technology, 2-12-1 Ookayama, Meguro-ku, Tokyo 152-8550, Japan}

\cftpagenumbersoff{figure}
\cftpagenumbersoff{table} 
\begin{document} 
\maketitle


\begin{abstract}
Understanding and reducing the in-orbit instrumental backgrounds are essential to achieving high sensitivity in hard X-ray astronomical observations. The observational data of the Hard X-ray Imager (HXI) on board the Hitomi satellite provides useful information on the background components, owing to its multi-layer configuration with different atomic numbers: the HXI consists of a stack of four layers of Si ($Z=14$) detectors and one layer of CdTe ($Z=48$, $52$) detector surrounded by well-type BGO (Bi$_{4}$Ge$_{3}$O$_{12}$) active shields. Based on the observational data, the backgrounds of top Si layer, the three underlying Si layers, and the CdTe layer are inferred to be dominated by different components, namely, low-energy electrons, albedo neutrons, and proton-induced radioactivation, respectively. Monte Carlo simulations of the in-orbit background of the HXI reproduce the observed background spectrum of each layer well, thereby verifying the above hypothesis quantitatively. In addition, {we suggest the inclusion of an electron shield to reduce the background}.
\end{abstract}


{\noindent \footnotesize\textbf{*}Kouichi Hagino,  \linkable{hagino@rs.tus.ac.jp} }

\begin{spacing}{1}

\section{Introduction}\label{sec:intro}
The Hard X-ray Imager (HXI) was one of the four observational instruments onboard the Hitomi satellite, which was launched on 2016 February 17\cite{Nakazawa2018,Takahashi2018}. Combined with the hard X-ray telescopes\cite{Tamura2018}, the HXI was designed to perform imaging spectroscopy in the 5--80~keV band. Although the Hitomi satellite was lost on 2016 March 25, the HXI was operational for 13 days in orbit, and provided useful insights on planning and designing future hard X-ray missions.

\begin{figure}[tbp]
\begin{center}
\includegraphics[width=0.4\hsize]{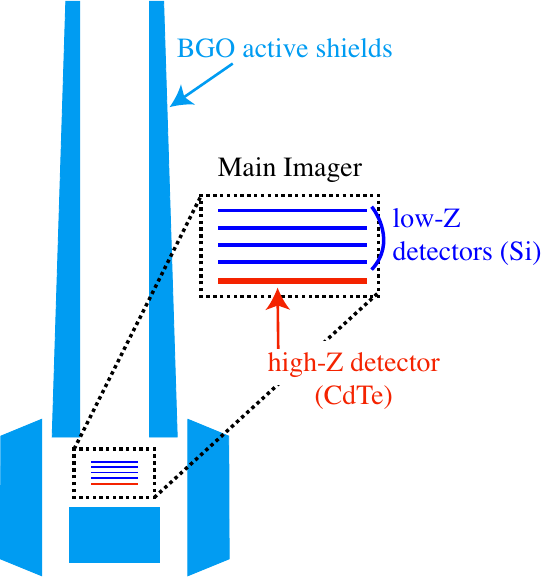}
\caption[Schematic structure of the HXI]{Schematic structure of the HXI. {This figure is adapted from Odaka~et~al.~2018\cite{Odaka2018}.}}
\label{fig:hxi}
\end{center}
\end{figure}

As shown in Fig.~\ref{fig:hxi}, the HXI consists of a stack of four layers of silicon (Si; $Z=14$) and one layer of cadmium telluride (CdTe; $Z=48, 52$) semiconductor detectors surrounded by well-type BGO (Bi$_{4}$Ge$_{3}$O$_{12}$) active shields\cite{Sato2016}. The BGO active shields have a typical thickness of $\sim 3{\rm ~cm}$, and their design is similar to that of the Hard X-ray Detector (HXD) onboard Suzaku\cite{Takahashi2007,Kokubun2007}, which efficiently reduced external backgrounds by the anti-coincidence technique. The main imager is composed of 0.5~mm thick Double-sided Si Strip Detector (DSSD) and 0.75~mm thick CdTe Double-sided Strip Detector (CdTe-DSD), with an imaging area of $32\times32{\rm ~cm^2}$. Four layers of DSSDs achieve detection efficiency of $>50$\% for X-rays below 30~keV. Combined with the CdTe-DSD, detection efficiency of $>70$\% is realized for X-rays up to 80~keV. The Si detectors detect lower-energy X-rays with less radioactivation background than high-$Z$ detectors while CdTe detectors detect higher-energy X-rays that cannot be absorbed by Si. This stacked configuration of the main imager is effective in optimizing the in-orbit instrumental background as well as in achieving high sensitivity in the hard X-ray band.

The multi-layer configuration of low-$Z$ (Si) and high-$Z$ (CdTe) sensors is very useful, not only in achieving high sensitivity, but also in decomposing the background components since interactions between the detector materials and cosmic-ray particles/photons have different Z-dependence. For example, the high-$Z$ materials are more sensitive to X/$\gamma$-ray photons than the low-$Z$ materials because photoabsorption cross-section is proportional to $Z^4$. On the other hand, high-$Z$ materials are less affected by neutrons because the recoil energy of nucleus via elastic scattering with neutron is roughly proportional to the inverse of the atomic mass number $A$. Therefore, the multi-layer configuration of the HXI would provide us fruitful knowledge on the origins of the in-orbit instrumental backgrounds, which is essential to achieve high sensitivity in hard X-ray astronomical observations.

In this paper, we describe our investigation of the origins of the HXI backgrounds using both observational data analysis and full Monte Carlo simulations. Section~\ref{sec:obs} describes the observational properties of the in-orbit instrumental backgrounds of the HXI. In Section~\ref{sec:sim}, we describe the Monte Carlo simulations of the HXI, and compare their results with the observational data. In Section~\ref{sec:force}, based on the results obtained from the simulations, we provide suggestions for future instruments in terms of reducing the in-orbit backgrounds, and finally, we provide the conclusions in Section~\ref{sec:conc}.

\section{Observational Properties of the HXI Background}\label{sec:obs}
\begin{figure}[tbp]
\begin{center}
\includegraphics[width=0.7\hsize]{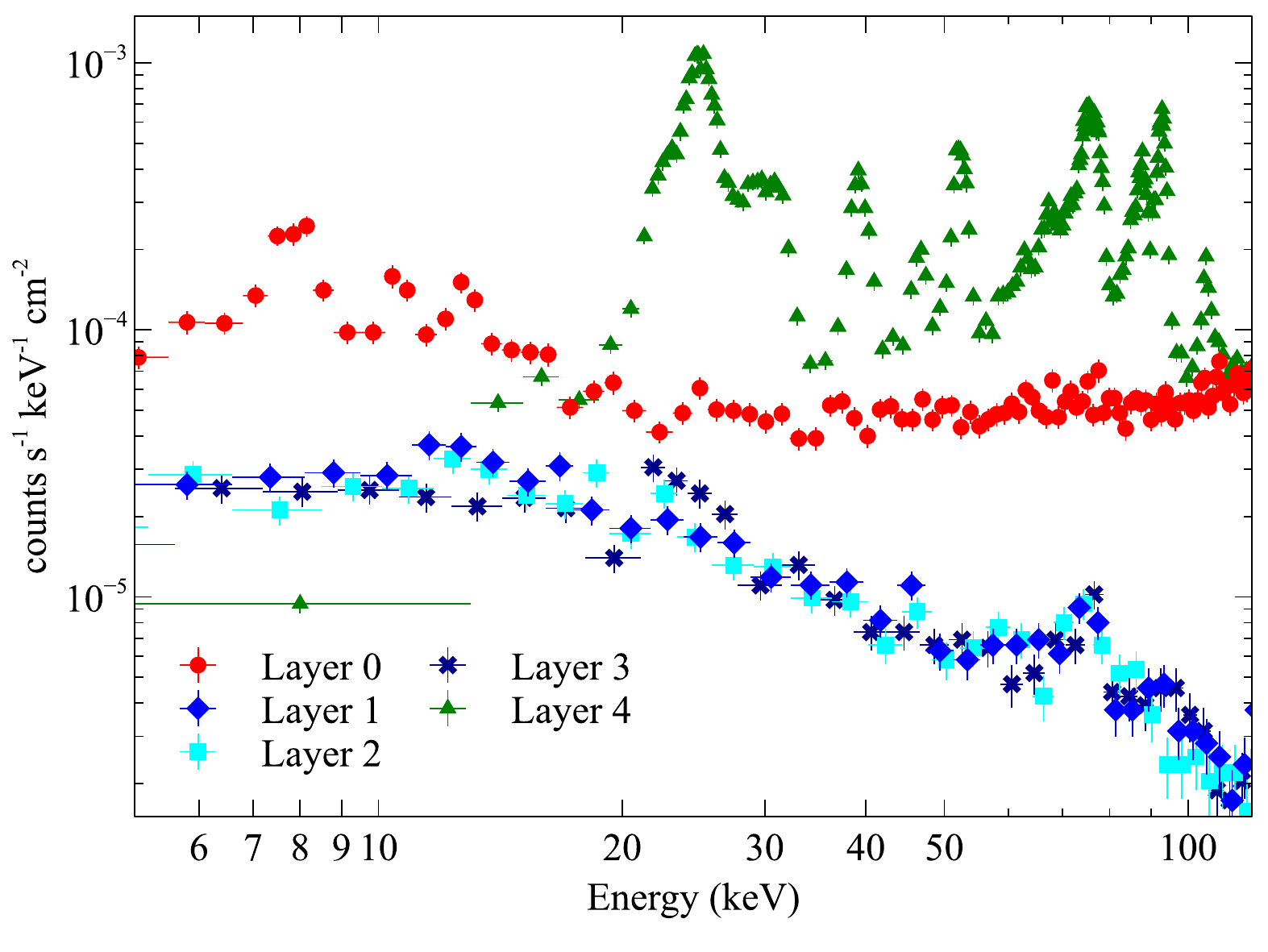}
\caption[Observed spectra of the instrumental background of the HXI]{Observed spectra of the instrumental background of the HXI, obtained from the Earth occultation data. Note that the vertical axis is per the area of the detector (not the mirror).}
\label{fig:obs}
\end{center}
\end{figure}

\begin{figure}[tbp]
\begin{center}
\includegraphics[width=0.7\hsize]{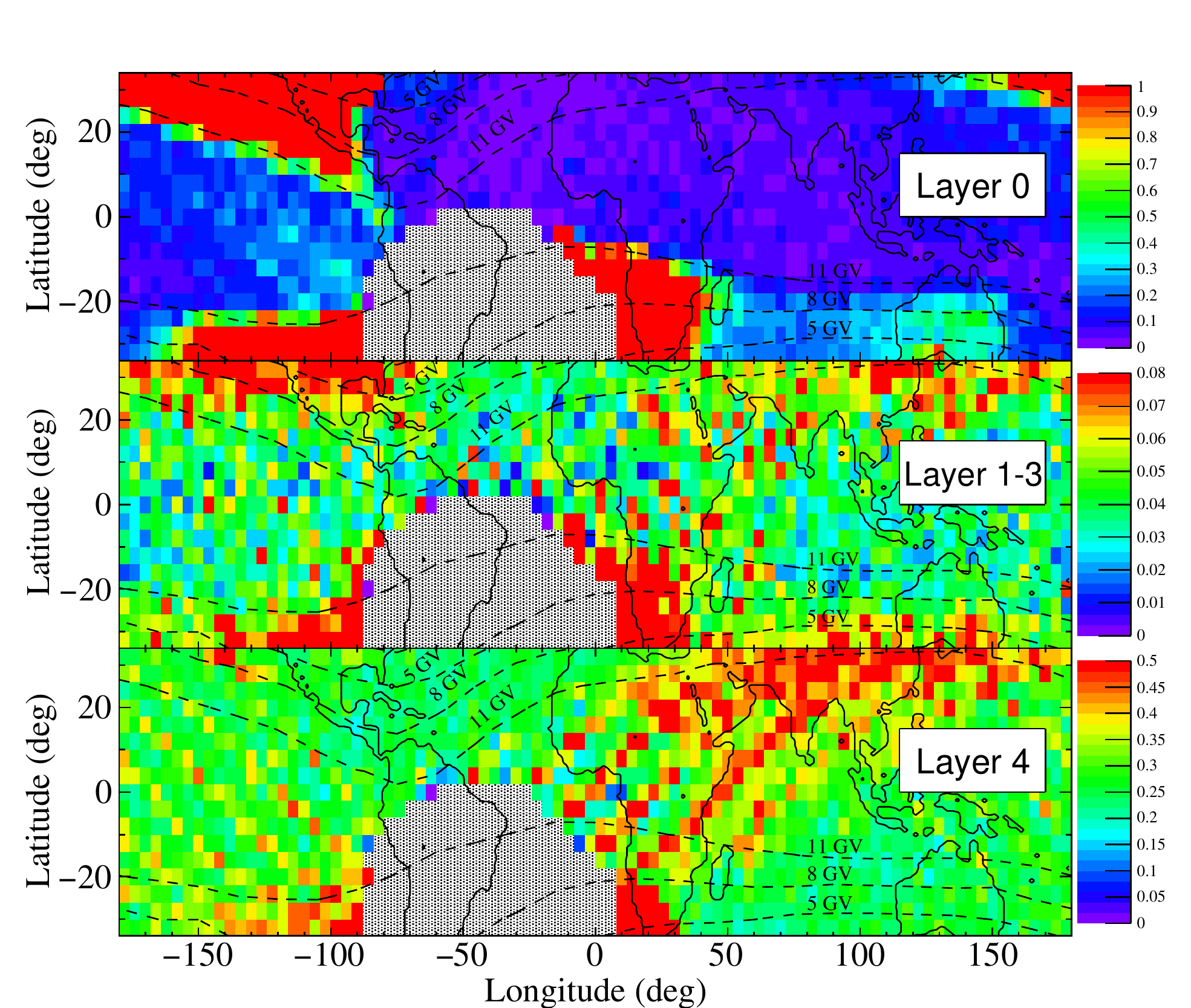}
\caption[Position dependence of the count rate of the top-layer DSSD, middle-layer DSSDs, and CdTe-DSD]{Position dependence of the count rate (${\rm counts~s^{-1}}$) of the top-layer DSSD (Layer~0; top panel), middle-layer DSSDs (Layer~1--3; middle panel), and CdTe-DSD (Layer~4; bottom panel) of the HXI during Earth occultation and blank sky observations. The black dashed lines indicate the geomagnetic cutoff rigidity. The gray region indicates the South Atlantic Anomaly where no data is downloaded from the satellite.}
\label{fig:saamap}
\end{center}
\end{figure}

Observed background spectrum of each layer of the HXI and its dependence on the satellite position (latitude/longitude) are shown in Fig.~\ref{fig:obs} and Fig.~\ref{fig:saamap}. Since the background spectra of Layer 1--3 (Fig.~\ref{fig:obs}) are similar to each other, hereafter, we separate the detectors of the HXI into three groups: the top layer of DSSD (Layer 0), remaining three layers of DSSD (Layer 1--3), and CdTe-DSD layer (Layer 4). In this section, we describe the observed properties of the backgrounds of each group and its origin suggested by the observations.

\subsection{Electron Background in the Top-layer DSSD}
The background spectrum of the top-layer DSSD (Layer~0; red circle in Fig.~\ref{fig:obs}) was a hard power law extending up to high energies of $\sim 100$~keV. In spite of its very large energy deposit ($\sim100$~keV), such a component was not seen in the other layers, indicating that the origin of this background component has a low penetrating power, while having energy of more than $\sim 100$~keV. This indicates that low-energy charged particles dominate the observed background.

In addition to the spectral properties, the variation of the count rate of this component with latitude and longitude of the satellite (top panel in Fig.~\ref{fig:saamap}) matches well with the distribution of the geomagnetically-trapped electrons measured by DEMETER/IDP\cite{Whittaker2013}. This distribution is clearly different from that of the geomagnetically-trapped protons, especially near the latitudes and longitudes corresponding to North America (latitude of $\gtrsim20^\circ$ and longitude of $\lesssim-100^\circ$). Therefore, the background of the top-layer DSSD is considered to originate from the geomagnetically-trapped low-energy electrons. 

\subsection{Neutron Background in the Middle-layer DSSDs}
The most remarkable property of the background spectra of the middle-layer DSSDs (Layer 1--3) is their similarity with each other. Such a property would be exhibited if the incident particle has a high penetrating power. Moreover, this component could not originate from the high-energy charged particles since the charged particles would be rejected because of anti-coincidence with the BGO active shields or with the other layers in onboard/on-ground data screening.

The similarity of the configuration of the detector (Si detector surrounded by well-type BGO active shields) and that present in Suzaku/HXD\cite{Takahashi2007,Kokubun2007} suggests that the background component of these detectors should also be similar. In addition, after scaling the HXD and HXI detectors with their corresponding thicknesses ({$0.2{\rm ~cm}$ and $0.05{\rm ~cm}$}, respectively), we find that the background of the HXD Si detector at 10--20~keV ({$\sim2\times10^{-4}{\rm ~counts~s^{-1}~keV^{-1}~cm^{-2}}/0.2{\rm ~cm}\simeq 1\times10^{-3}{\rm ~counts~s^{-1}~keV^{-1}~cm^{-3}}$}) is consistent with that of the middle-layer DSSDs in HXI ({$\sim3\times10^{-5}{\rm ~counts~s^{-1}~keV^{-1}~cm^{-2}}/0.05{\rm ~cm}\simeq0.6\times10^{-3}{\rm ~counts~s^{-1}~keV^{-1}~cm^{-3}}$}) within a factor of two. In the case of the HXD Si detector, the background is thought to be dominated by albedo neutrons\cite{Fukazawa2009,Mizuno2010}. One of the supporting evidences is that it shows an anti-correlation with the cut-off rigidity. {In other words, the background of HXD Si detector is higher at the locations with lower cut-off rigidity. Since the cut-off rigidity is a parameter describing the geomagnetic shielding from the cosmic-ray particles, the lower cut-off rigidity means that there are a lot of cosmic-ray particles and their secondary particles such as albedo neutrons.} As shown in the middle panel of Fig.~\ref{fig:saamap}, the middle-layer DSSDs in HXI also have such a property. Therefore, similarly to the HXD Si detector, the background of the middle-layer DSSDs in HXI should also be dominated by albedo neutrons.

\subsection{Proton-induced Radioactivation Background in the CdTe-DSD}
Unlike the feature-less spectrum of DSSDs, the background spectrum of CdTe-DSD (Layer 4; green triangle in Fig.~\ref{fig:obs}) is composed of many emission lines. Most of these lines originate from the radioactive isotopes arising from the activation of Cd and Te, induced by geomagnetically-trapped protons in the South Atlantic Anomaly (SAA). Although this effect is inevitable, it is mitigated by the thick BGO shields with a typical thickness of $\sim3{\rm ~cm}$, which stop protons with energies $\lesssim 100{\rm ~MeV}$.

The background of irradiated CdTe-DSD was thoroughly investigated with Monte Carlo simulations in our previous work\cite{Odaka2018}. In that paper, we successfully reproduced the observed background of CdTe-DSD with the simulations, and revealed that it is dominated by the proton-induced radioactivation of the CdTe-DSD itself.

\section{Monte Carlo Simulations of the HXI Background}\label{sec:sim}
In order to verify the hypothesis of the origins of the HXI backgrounds in the previous section, we performed a full Monte Carlo simulation by taking account of the in-orbit radiation environment. We used a Monte Carlo simulation framework {\sc ComptonSoft}\cite{Odaka2010}, identical to that used in a previous work on the radioactivation background of the CdTe-DSD\cite{Odaka2018}. In the {\sc ComptonSoft}, interactions between incident particles and detector materials are simulated by the software based on the {\sc Geant4} toolkit library\cite{Agostinelli2003,Allison2006,Allison2016}. In addition, detailed detector responses due to the charge carrier transport inside semiconductor detectors are also implemented. In this work, we used the same parameter values for the detector response as those tuned to reproduce the experimental data\cite{Hagino2018}. Most of the passive materials as well as the imager module and active shields are implemented in the simulation to accurately calculate interactions in the radiation environment. 

\begin{figure}[tbp]
\begin{center}
\includegraphics[width=0.7\hsize]{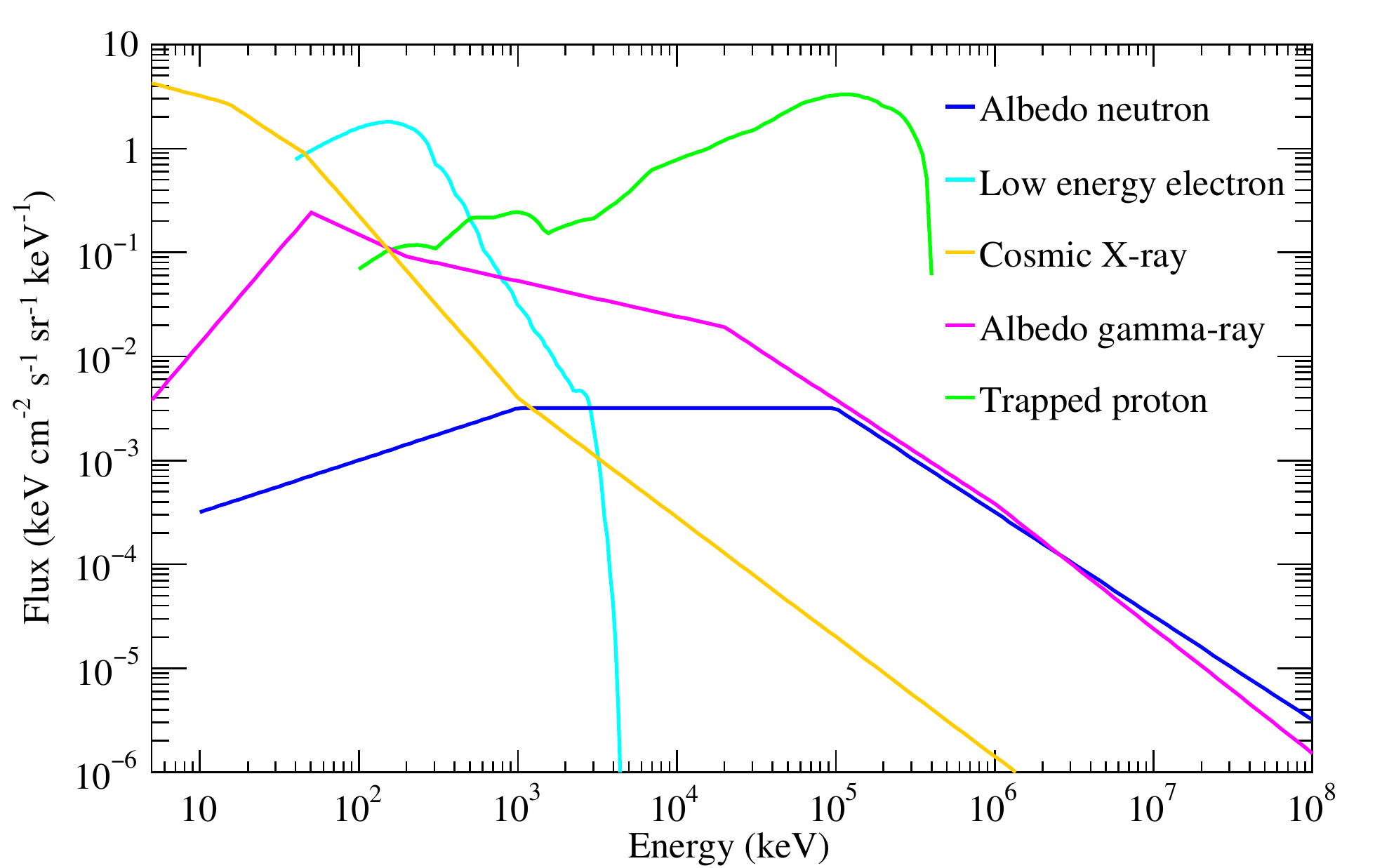}
\caption[Radiation environment spectral model of the Monte Carlo simulations]{Radiation environment spectral model of the Monte Carlo simulations.}
\label{fig:rad}
\end{center}
\end{figure}

\subsection{Radiation Environment}
The radiation environment assumed in this work is shown in Fig.~\ref{fig:rad}. It is composed of albedo neutrons, geomagnetically trapped protons, low-energy electrons, cosmic X-ray and albedo gamma-ray. The flux and spectral shapes of albedo neutrons, cosmic X-ray and albedo gamma-ray were based on our previous study \cite{Mizuno2010}. Geomagnetically trapped protons were obtained from Space Environment Information System (SPENVIS)\cite{SPENVIS} AP-8 model at solar minimum.

The spectral shape of the low-energy electrons was also obtained from SPENVIS AE-8 model, but its flux was estimated from the observed trigger rate of the top-layer DSSD. Since we eliminated the SAA and the other high flux regions in the analysis (see Hagino~et~al.,~2018\cite{Hagino2018} for more details), the electron flux averaged over all the orbital phase obtained from AE-8 is an overestimation. Although the orbital dependence of the electron flux is implemented in AE-8 model, the lower flux region included in our analysis is not fully reproduced in the model. Thus, we scale the AE-8 flux by a ratio of the observed trigger rate in the low flux region to the orbit-averaged trigger rate. The scaling factor is estimated to be $4.2\times 10^{-4}\textrm{--}6.7\times 10^{-4}$. In the trapped electron spectrum in Fig.~\ref{fig:rad}, the upper limit of the scaling factor is adopted because it gives a simulated background spectra that matches well with the observed spectrum. It is worth noting that these radiation models except for the low-energy electrons were not optimized for the observed data of the HXI.

\subsection{Comparison with the Observed Spectra}
\begin{figure}[tbp]
\begin{center}
\includegraphics[width=\hsize]{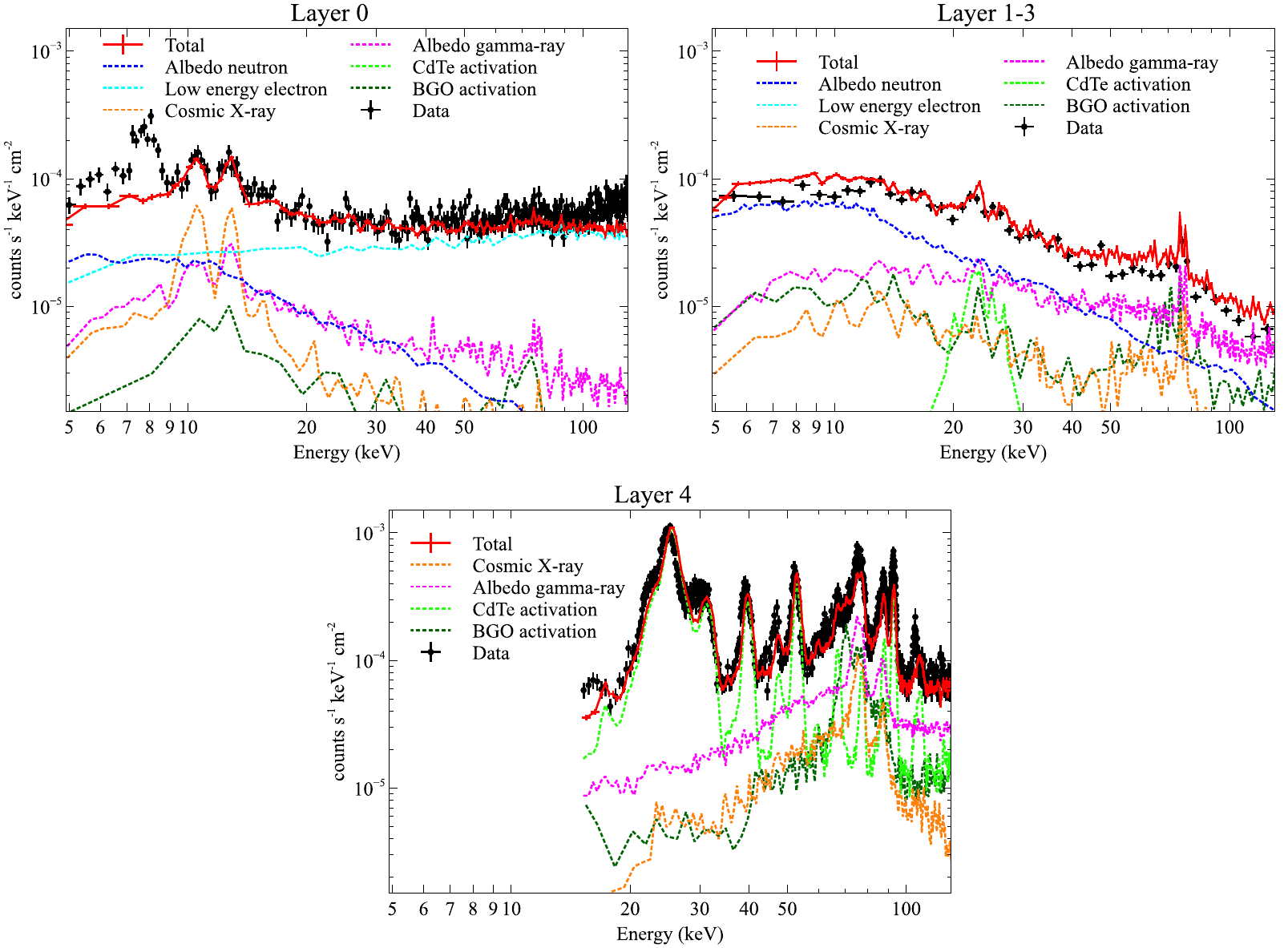}
\caption[Simulated background spectra of the top-layer DSSD, middle-layer DSSDs and CdTe-DSD]{Simulated background spectra of top-layer DSSD (Layer 0), middle-layer DSSDs (Layer 1--3) and CdTe-DSD (Layer 4) of the HXI (red solid lines), compared with the observed data (black points). The dashed lines represent the contributions from each components. Albedo neutrons and low-energy electrons are not shown in the bottom panel (Layer 4) because their contribution is negligible.}
\label{fig:sim}
\end{center}
\end{figure}
The simulated background spectrum of each layer is shown in Fig.~\ref{fig:sim}. The observed background spectra are well reproduced by the Monte Carlo simulations with albedo neutrons, low-energy electrons, trapped protons, cosmic X-rays, albedo gamma rays, and radioactivation of CdTe and BGO. In all the layers, the difference between the simulation results and observations is less than a factor of two, except for the line-like feature at 8~keV in the top-layer DSSD (Layer 0). Although the origin of this small excess is unclear, it is possibly due to the fluorescence X-ray of copper at 8.0~keV induced by the radiation environment in space. These X-rays might be emitted from materials that are not implemented in the simulation, such as the wire harness and/or the extensible optical bench. The other two lines at 10-12~keV in the top-layer DSSD are the fluorescence X-rays from the CXB shield made of lead.

According to the simulations, the background spectra of the top-layer DSSD (Layer 0), middle-layer DSSDs (Layer 1--3), and CdTe-DSD (Layer 4) are dominated by low-energy electrons, albedo neutrons, and proton-induced radioactivation, respectively. These results confirm the validity of the hypothesis based on the spectral features of observational data analysis in Sec.~\ref{sec:obs}.

\section{Suppression of Low-Energy Electron Background}\label{sec:force}
As shown in Fig.~\ref{fig:obs}, the background of the middle-layer DSSDs (Layer 1--3) has been successfully reduced down to $\sim3\times10^{-5}{\rm ~counts~s^{-1}~keV^{-1}~cm^{-2}}$, so that the remaining background is almost only due to the albedo neutron component, which is difficult to reduce. However, the background of the top-layer DSSD (Layer 0) is dominated by the electron component, which degrades the sensitivity to below $20$~keV where DSSD is most sensitive. Therefore, reducing this electron component is essential to achieve better sensitivity in future missions. In this section, we discuss the background that results from low-energy electrons, and propose a possible design to reduce it.

\subsection{Incident Path of Low Energy Electrons}
\begin{figure}[tbp]
\begin{center}
\includegraphics[width=\hsize]{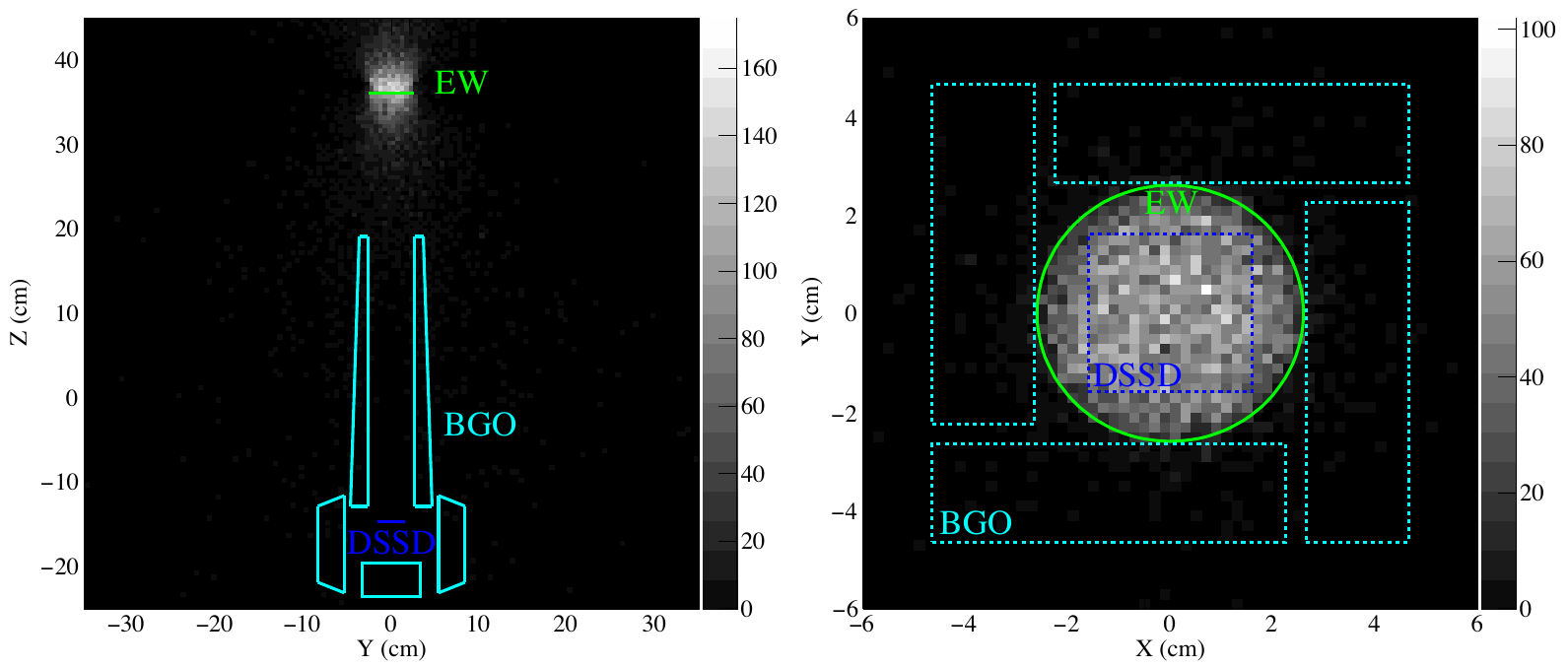}
\caption[Incident path of the low-energy electrons]{Incident path of the low-energy electrons. {\it Left}: Intersections of the initial electron paths and a vertical plane ($X=0$). {\it Right}: Intersections of the initial electron paths and a horizontal plane at the height of the entrance window (EW; $Z=36.1$~cm).}
\label{fig:path}
\end{center}
\end{figure}

In Section~\ref{sec:obs} and \ref{sec:sim}, we showed that the low-energy electrons with energies of $\sim 100{\rm ~keV}$ are the major background component of the top-layer DSSD. However, electrons with energy of $100{\rm ~keV}$ are easily stopped by silicon with a thickness of $100{\rm ~\mu m}$. These electrons cannot enter the DSSD directly because the HXI is designed to block all such direct paths. Therefore, in order to reduce the electron component, it is necessary to understand the incident path of the low-energy electrons.

The left panel of Fig.~\ref{fig:path} shows the intersections of a vertical plane crossing the detector center ($X=0$) and the initial path of the low-energy electrons entering the top-layer DSSD. This plot shows the path of the low-energy electrons on the vertical plane $X=0$, assuming that the electrons do not change their direction before reaching this plane. From this figure, we can infer that most of the electrons hit the entrance window (EW; two layers of 30-$\mu$m-thick poly-carbonate sheets). It becomes more apparent if we see the intersections of the electron paths with a horizontal plane placed at the height of the EW (right panel of Fig.~\ref{fig:path}). According to this plot, almost all incident paths intersect with the EW. It indicates that most of the low-energy electrons are scattered at the EW, and then enter the top-layer DSSD. Since the EW needs to be thin to avoid absorbing X-rays at $5{\rm ~keV}$, it is impossible to block the low-energy electrons at the EW. Therefore, in order to reduce the electron background, it is important to reduce the number of electrons hitting the EW.

\subsection{{Cutting Electron Component}}
In order to reduce the low-energy electron background, the EW should be shielded from the low-energy electrons. The shield should have a density thickness more than $\sim 10^{-2}{\rm ~g~cm^{-2}}$, corresponding to the CSDA (continuous slowing down approximation) range of 100~keV electrons. If carbon fiber reinforced polymer (CFRP; density $\sim1.5{\rm ~g~cm^{-3}}$) is chosen as the electron shield, the required thickness is less than $\sim 100{\rm ~\mu m}$. It is worth mentioning that adding such shield will not increase the instrumental background since the material from which it is made does not activate. Therefore, this requirement is relatively easy to satisfy.

\begin{figure}[tbp]
\begin{center}
\includegraphics[width=0.8\hsize]{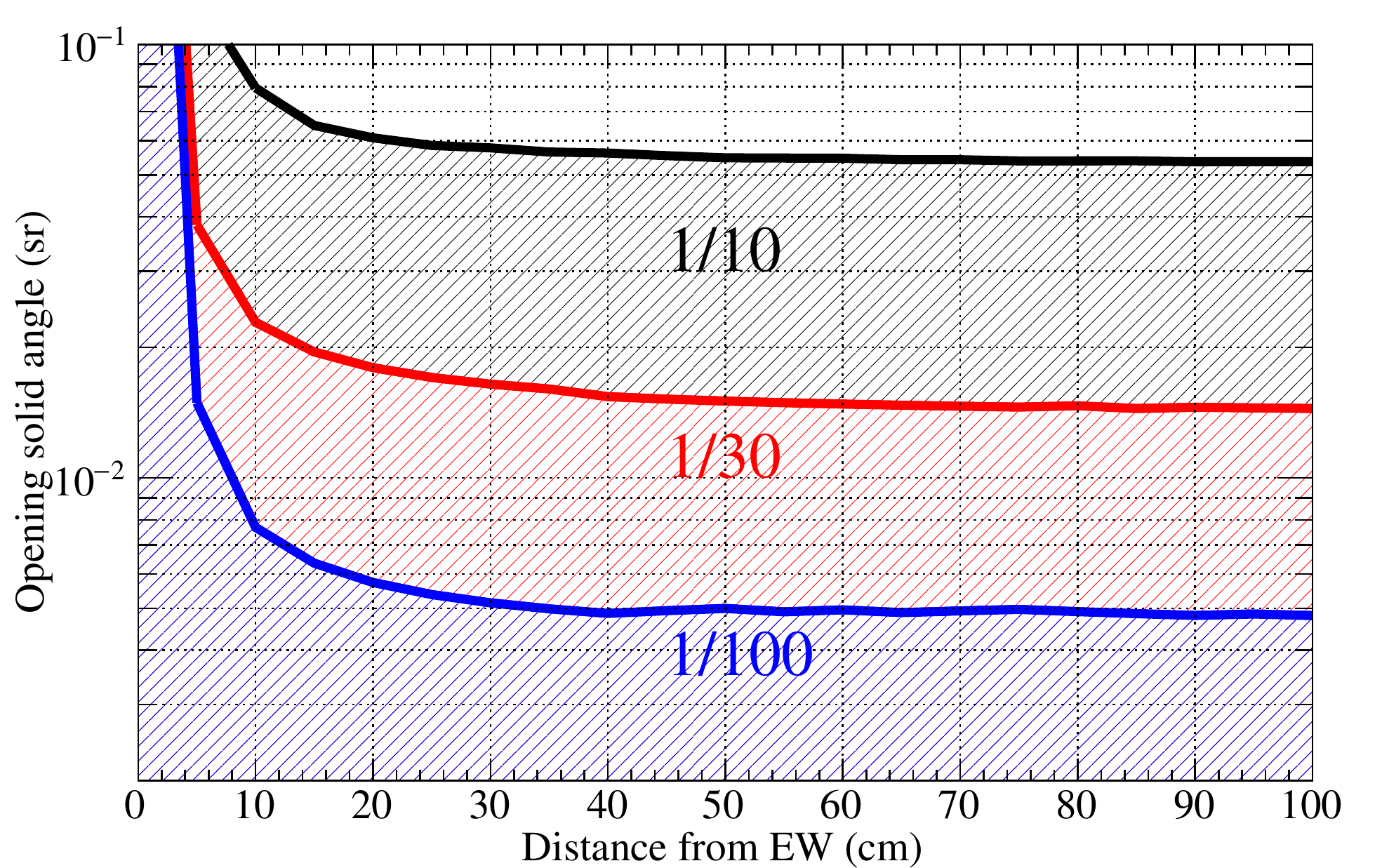}
\caption[Opening solid angle of the shield]{Opening solid angle of the shield required to reduce the electron background by 1/10, 1/30, and 1/100. The required solid angle depends on the distance from the EW because the size of the EW is not negligible at small distances.}
\label{fig:shield}
\end{center}
\end{figure}

In addition to the thickness, the electron shield must cover a large solid angle of the sky when seen from the EW. Fig.~\ref{fig:shield} shows the requirement of an opening solid angle of the shield calculated from the simulation data. In this calculation, the shield is assumed to have a circular hole just above the EW. The horizontal axis is the distance between the hole and EW. According to this figure, the solid angle of the hole seen from the EW should be smaller than $0.05$, $0.01$, and $0.005{\rm ~sr}$ to reduce the electron background by $1/10$, $1/30$, and $1/100$, respectively.

\begin{figure}[tbp]
\begin{center}
\includegraphics[width=0.7\hsize]{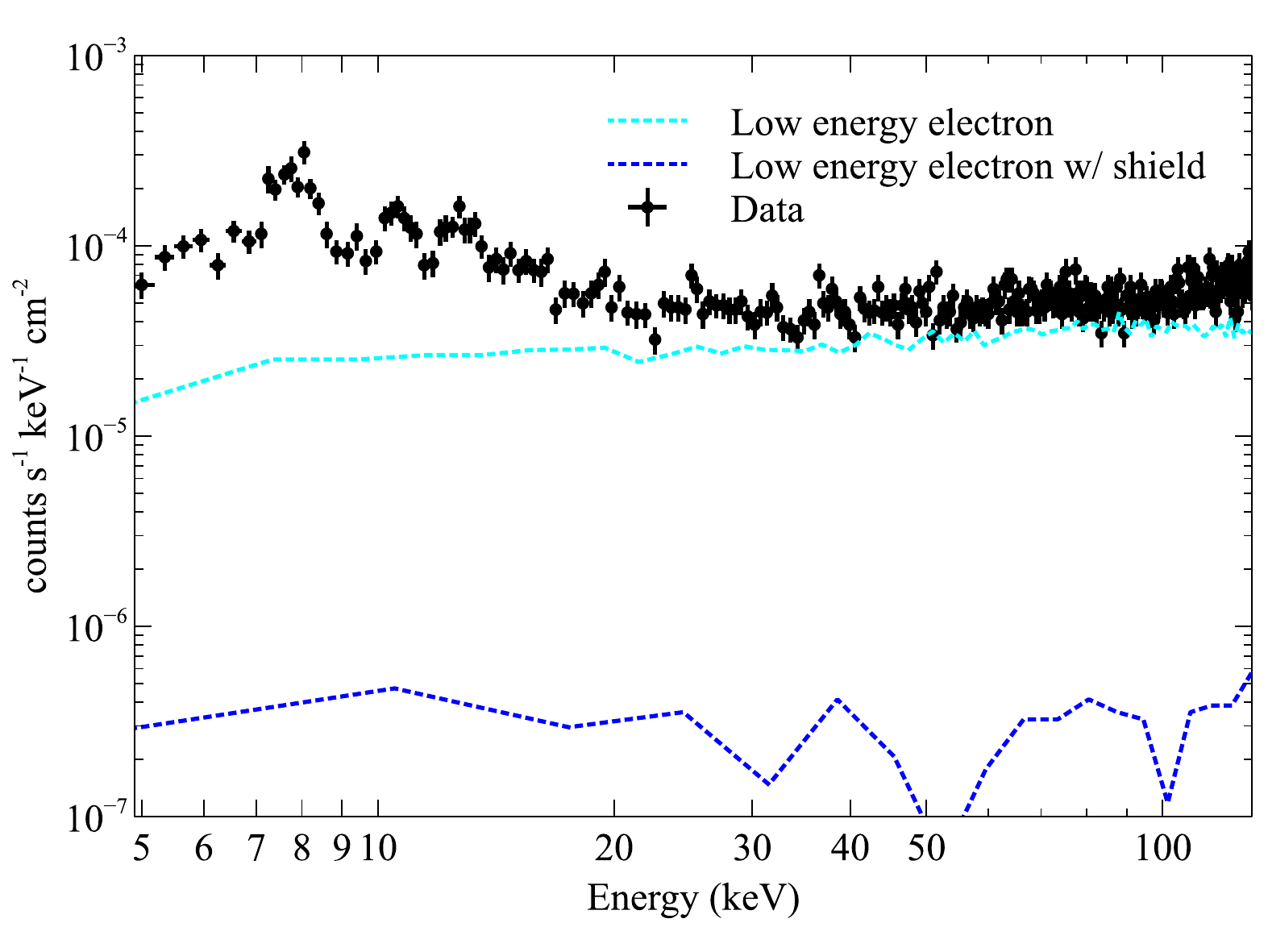}
\caption[Simulated electron background spectrum of the top-layer DSSD with shield]{Simulated electron background spectrum of the top-layer DSSD with shield. With an appropriate shield, the electron background can be reduced. The large fluctuation in the simulated spectrum with shield is due to the Monte Carlo noise.}
\label{fig:spec_shield}
\end{center}
\end{figure}

As a demonstration, we show the electron background spectrum of the top-layer DSSD with an appropriate shield in Fig.~\ref{fig:spec_shield}. The shield is assumed to cover almost all of the solid angle except for a circular hole with a diameter of $10{\rm ~cm}$ at $1{\rm ~m}$ above the EW, corresponding to the opening solid angle of $\sim0.008$. Here, all the electrons coming from the solid angle covered by the shield are assumed to be completely blocked. Since such a shield can efficiently reduce the number of low-energy electrons from outside of the field of view, the electron background spectrum with the shield is nearly two orders of magnitude lower than that without the shield. Hence, as demonstrated, the electron background can be reduced in future missions if an appropriate shielding is applied.

{Although we showed that the electron background can be reduced by adding an appropriate shield to the HXI, this design is not necessarily the best one for all hard X-ray missions. The best design must depend on the scientific objectives and constraints on the satellite design. When designing a future mission, a detailed quantitative analysis of the relative merits and demerits of the various possible detector designs would be needed.}

\section{Conclusions}\label{sec:conc}
The observational data of the HXI indicates the origins of their instrumental backgrounds: the backgrounds of top-layer Si, middle-layer Si and bottom-layer CdTe are inferred to originate from low-energy electrons, albedo neutrons and proton-induced radioactivation, respectively. We successfully reproduced the spectral shape and flux of the observed HXI backgrounds using the full Monte Carlo simulations. Based on the results from this simulation, we conclude that the electron background is expected to be reduced after applying appropriate shielding, which can further benefit future missions.

\appendix    

\acknowledgments 
We acknowledge all the {\it Hitomi} team members, including many graduate students, for their great contributions to the HXI and the {\it Hitomi} project. We acknowledge the support of JSPS/MEXT KAKENHI grant numbers 24105007, 15H03639, 25287059, 24244014, 16H02170, and the JSPS Core-to-Core Program. All the members from the U.S. acknowledge support received from the NASA Science Mission Directorate. Stanford and SLAC members acknowledge support via DoE contract to SLAC National Accelerator Laboratory DE-AC3-76SF00515 and NASA grant NNX15AM19G. French members acknowledge support from the Centre National d'Etudes Spatiales (CNES).


\bibliography{report}   
\bibliographystyle{spiejour}   


\vspace{2ex}\noindent\textbf{Kouichi Hagino} is an assistant professor at Tokyo University of Science. He received his BS and MS degrees in physics from the University of Tokyo in 2010 and 2012, respectively, and his PhD degree in physics from the University of Tokyo in 2015. He has been working in development of semiconductor detectors for applications in high-energy astrophysics.

\vspace{1ex}

\listoffigures

\end{spacing}
\end{document}